\documentclass{article}
\usepackage{spconf}
\usepackage[caption=false,font=normalsize,labelfont=sf,textfont=sf]{subfig}

\usepackage{cite}
  \usepackage[pdftex]{graphicx}
  \graphicspath{{./figures/}}
\usepackage{amsmath,amsfonts,amssymb}
\usepackage{algorithm}
\usepackage{algorithmicx}
\usepackage{algpseudocode}
\usepackage{fixltx2e}
\usepackage{stfloats}
\usepackage{url}
\usepackage{multirow}
\usepackage{color}
\usepackage{acro}

\newcommand{\azi}{y^\text{AZI}}
\newcommand{\ele}{y^\text{ELE}}
\newcommand{\eazi}{e_\text{AZI}}
\newcommand{\eele}{e_\text{ELE}}
\newcommand{\sazi}{\sigma_\text{AZI}}
\newcommand{\sele}{\sigma_\text{ELE}}
\newcommand{\VMF}{\mathrm{VMF}}
\newcommand{\atan}{\mathrm{atan}}
\newcommand{\pos}{\theta}

\newcommand{\eye}{I}

\newcommand{\zeros}{\mathrm{O}}

\renewcommand{\d}{\mathop{}\!\mathrm{d}}
\newcommand{\N}{\mathrm{N}}

\renewcommand{\t}{\mathsf{T}}

\newcommand{\np}{N_\text{p}}

\DeclareAcronym{aoa}{
	short=AOA,
	long=angle-of-arrival
}
\DeclareAcronym{pf}{
	short=PF,
	long=particle filter
}
\DeclareAcronym{kf}{
	short=KF,
	long=Kalman fiter
}
\DeclareAcronym{ekf}{
	short=EKF,
	long=extended Kalman fiter
}
\DeclareAcronym{ukf}{
	short=UKF,
	long=unscented Kalman filter
}
\DeclareAcronym{vmf}{
	short=VMF,
	long=von Mises--Fisher
}
\DeclareAcronym{pdf}{
	short=PDF,
	long=probability density function
}
\DeclareAcronym{rmse}{
	short=RMSE,
	long=root-mean-square error
}
\begin{document}

\title{3D ANGLE-OF-ARRIVAL POSITIONING USING VON MISES--FISHER DISTRIBUTION}

\name{Henri Nurminen\thanks{Henri Nurminen has received funding from Tampere University of Technology Graduate School, Nokia Technologies Oy, the Foundation of Nokia Corporation, Tekniikan edist\"amiss\"a\"ati\"o, and Emil Aaltonen Foundation.}, Laura Suomalainen, Simo Ali-L\"oytty, and Robert Pich\'e}
\address{Tampere University of Technology, Tampere, Finland}

\maketitle

\begin{abstract}
We propose modeling an \ac{aoa} positioning measurement as a \ac{vmf} distributed unit vector instead of the conventional normally distributed azimuth and elevation measurements. Describing the 2-dimensional \ac{aoa} measurement with three numbers removes discontinuities and reduces nonlinearity at the poles of the azimuth--elevation coordinate system. Our computer simulations show that the proposed \ac{vmf} measurement noise model based filters outperform the normal distribution based algorithms in accuracy in a scenario where close-to-pole measurements occur frequently.
\end{abstract}
\begin{keywords}
angle-of-arrival; positioning; von Mises--Fisher distribution; particle filter; extended Kalman filter
\end{keywords}

\section{Introduction} \acresetall \label{sec:introduction}

Many future positioning systems will use \ac{aoa} measurements, as the coming 5G networks can be equipped with antenna arrays that enable measuring the \ac{aoa} of the received electromagnetic signal \cite{koivisto2017a}. An \ac{aoa} measurement consists of two components: azimuth and elevation. A conventional approach is to model the measurements as noisy versions of the true azimuth and elevation \cite{toloei2014,sofyali2015,huang2016,ahmed2016,koivisto2017b}, and use \ac{ekf} or \ac{ukf} that assume that the measurement noises of azimuth and elevation follow normal distributions. However, this model is problematic in a number of ways: I In this model the solid angle of measurement uncertainty is smaller close to the ``pole'' directions, i.e.\ the two directions where azimuth is not defined. II The measurement model is highly nonlinear close to the poles and discontinuous in the pole, which makes gradient-based approximations for optimisation and extended Kalman filtering unstable. 3) Rotations of the spherical coordinate system in which the azimuth and elevation are expressed change the measurement error model. The nonlinearity problem has been reported to result in divergence of the \ac{ekf} \cite{sofyali2015}.

To remedy these problems, we propose expressing the 2-dimensional spherical \ac{aoa} measurement as a 3-dimensional Cartesian unit vector, and modeling the measurement error with the \ac{vmf} distribution \cite{fisher1953,fisher1987}. This idea and its advantages are analogous to modeling a rotation with a 4-dimensional Bingham-distributed unit quaternion instead of a 3-dimensional Euler angle set \cite[Ch.\ 3.10]{kuipers1999}, \cite{gilitschenski2016}. We also propose a \ac{pf} algorithm \cite{gordon1993} based on the \ac{vmf} measurement error mode, and \ac{ekf} \cite[Ch.\ 8.3]{jazwinski1970} and \ac{ukf} \cite{julier1995} algorithms that approximate the \ac{vmf} update with the assumption that the unit vector measurement is the true direction's unit vector plus a trivariate normal noise. Our simulations show that the proposed positioning algorithms outperform the conventional algorithms in accuracy. \ac{vmf} filters have also been proposed in \cite{chiuso1998, traa2014, markovic2014, kurz2016}, but in these filters both the state and measurements are \ac{vmf}-distributed unit vectors.

\section{Modelling of AOA measurement}

\subsection{Von Mises--Fisher distribution} \label{sec:distribution}

The support of the \ac{vmf}'s \ac{pdf} is the unit (hyper-)sphere. A unit vector $x \!\in\! \mathbb{R}^n$ follows the distribution $\VMF(\mu,\kappa)$ with mean direction $\mu \!\in\! \mathbb{R}^n$, $\|\mu\|\!=\!1$, and concentration parameter $\kappa \!\in\! \mathbb{R}_+$ if its \ac{pdf} is
\begin{equation} \label{eq:vmf_prob} 
p(x)= C_\kappa e^{\kappa \mu^\t x}.
\end{equation}
The larger the parameter $\kappa$ is, the more the probability mass is concentrated around the direction $\mu$. For $\kappa\!>\!0$ the distribution is unimodal and for $\kappa\!=\!0$ it is uniform on the sphere. For a 3-dimensional variable the normalisation constant is
\begin{equation}
C_\kappa = \left\{ \begin{array}{ll}\tfrac{\kappa}{4\pi\sinh\kappa},&\kappa>0\\[0.5ex]\tfrac{1}{4\pi},&\kappa=0 \end{array} \right..
\end{equation}

The \ac{vmf} distribution is suitable for modelling directional data because a direction can be bijectively mapped to a unit vector. The distribution is rotation invariant in the sense that if $x\!\sim\!\VMF(\mu,\kappa)$, then for $y\!=\!Rx$ holds $y\!\sim\!\VMF(R\mu,\kappa)$ for a rotation matrix $R$. The \ac{pdf} of $\VMF(\mu,\kappa)$ is the restriction of the \ac{pdf} of the multivariate normal distribution $\N(\mu,\tfrac{1}{\kappa} \eye_n)$ into the origin-centered unit hyper-sphere \cite[Ch.\ 9.3.2]{mardia2000}.

\subsection{Comparison of normal and \ac{vmf} models}

In this paper an \ac{aoa} measurement consists of azimuth measurement $\azi \!\in\! (-\pi,\pi]$ and elevation measurement $\ele \!\in\! [-\tfrac{\pi}{2},\tfrac{\pi}{2}]$. We define the equator to be $\ele\!=\!0$, the poles $\ele\!=\!\pm\tfrac{\pi}{2}$, and the up direction $\ele\!=\!\tfrac{\pi}{2}$. The mapping from an \ac{aoa} measurement to a unit vector is thus
\begin{equation}
\mathtt{TO\_UNITVECTOR}(\azi,\ele) = \left[\begin{smallmatrix} \cos\azi\cdot\cos\ele \\ \sin\azi\cdot\cos\ele \\ \sin\ele \end{smallmatrix}\right].
\end{equation}
Given user position $\theta \!\in\! \mathbb{R}^3$ and anchor position $s \!\in\! \mathbb{R}^3$, the conventional normal distribution based measurement model is
\vspace{-3mm}
\begin{subequations} \label{eq:normal_model}
\begin{align}
&\azi = \atan_2(\pos_2-s_2, \pos_1-s_1) + \eazi\\
&\ele = \atan_2((\pos_3-s_3), \|\pos_{1:2}-s_{1:2}\|) + \eele,
\end{align}
\end{subequations}
where $\eazi\!\sim\!\N(0,\sazi^2)$ and $\eele\!\sim\!\N(0,\sele^2)$ are noise terms that are statistically mutually independent and independent from $\pos$, and $\sazi$ and $\sele$ are model parameters. The proposed \ac{vmf} based measurement model is
\begin{equation} \label{eq:vmf_model}
\mathtt{TO\_UNITVECTOR}(\azi,\ele) \sim \VMF\big(\tfrac{\pos-s}{\|\pos-s\|}, \kappa\big),
\end{equation}
where the concentration $\kappa$ is a model parameter.

In order to compare the normal and \ac{vmf} based estimation algorithms, we seek a simple rule-of-thumb formula that converts one model to the other. When $\alpha_{x,\mu}$ is the angle between unit vectors $x$ and $\mu$, the \ac{pdf} of $x\!\sim\!\VMF(\mu,\kappa)$  is
\begin{align}
p(x) &\propto e^{\kappa \cos(\alpha_{x,\mu})} \approx e^{\kappa (1-\frac{1}{2}\alpha_{x,\mu}^2)} \label{eq:cos_taylor} \propto \N(\alpha_{x,\mu}; 0, \tfrac{1}{\kappa}),
\end{align}
which follows from the second order truncated MacLaurin series of $\cos(\alpha_{x,\mu})$ and holds for small $\alpha_{x,\mu}$. We thus recommend to implement the normal distribution based filters for \ac{vmf}-distributed errors and \ac{vmf} based filters for normally distributed errors with the conversion rules
\begin{equation} \label{eq:sigma2kappa}
[\sazi^2]_\text{filter} \!=\! [\sele^2]_\text{filter} \!\triangleq\! \tfrac{1}{\kappa_\text{true}},\ \kappa_\text{filter} \!\triangleq\! \tfrac{1}{\max\{[\sazi^2]_\text{true},[\sele^2]_\text{true}\}}.
\end{equation}

\section{Bayesian filtering} \label{sec:filters}

We assume a normal initial prior $x_0\!\sim\!\N(x_{0|0},P_{0|0})$ and a linear--normal state transition model for the state $x \in \mathbb{R}^{n_x}$
\begin{equation}
x_k = A_{k-1} x_{k-1} + w_{k-1},\quad w_{k-1} \sim \N(0,Q_{k-1}),
\end{equation}
where $A_{k-1}$ is state transition matrix, $w_{k-1}$ is process noise, and $Q_{k-1}$ is process noise covariance matrix. In this paper the state includes the 3-dimensional user position, and the three position components in the state are denoted by $[x_k]_\text{pos}$.

In the \ac{pf} algorithm \cite[Ch.\ 3]{ristic2004} random samples (``particles'') are generated from the initial prior, propagated in time using the state transition model, weighted using the measurement information, and resampled when the weight concentrates too much. \ac{pf} is flexible in modeling and can be applied to both normal distribution based measurement model \eqref{eq:normal_model} and \ac{vmf} model \eqref{eq:vmf_model} without any application-specific tweaks. \ac{pf} for the \ac{vmf} model is given in Algorithm \ref{algo:pf}.

\Ac{ekf} and \ac{ukf} are nonlinear Kalman filter extensions for state-space models where the noises are normally distributed but the model functions can be nonlinear. Application of \ac{ekf} and \ac{ukf} to the normal model \eqref{eq:normal_model} is straightforward, except that the angle wrappings have to be taken into account when computing the angular differences. Because \ac{ekf} and \ac{ukf} assume normally distributed measurement noise, they are not applicable to the \ac{vmf} measurement model \eqref{eq:vmf_model}, but we approximate the \ac{vmf} model with 
\begin{equation} \label{eq:approx_vmf_model}
\mathtt{TO\_UNITVECTOR}([\azi_k]_j,[\ele_k]_j) \sim \N\big(\tfrac{[x_k]_\text{pos}-s_j}{\| [x_k]_\text{pos}-s_j \|}, \tfrac{1}{\kappa} \eye_3 \big).
\end{equation}
The details of the computation of the measurement model function and its Jacobian are given in Algorithm \ref{algo:meas_func}.

\begin{algorithm}[t]
\small
\caption{Particle filter for VMF measurement noise}\label{algo:pf}
\begin{algorithmic}[1]
\newcommand{\dl}{\delta}
\State \textbf{Inputs:} initial prior $x_{0|0}$, $P_{0|0}$; state-transition model $A_{1:K}$, $Q_{1:K}$; concentration parameter $\kappa$; $\azi_{1:K}$, $\ele_{1:K}$; anchor positions \& rotations $s_{1:n_s}$, $R_{1:n_s}$
\State \textbf{Outputs:} estimates $x_{k|k}$ for $k\!=\!1,\ldots,K$
\State $x_0^{(i)} \sim \N(x_{0|0},P_{0|0})$,\quad$w_0^{(i)} \gets \frac{1}{\np}$ for $i\!=\!1,\ldots,\np$
\For{$k=1:K$}
	\State $u_j \!\!\gets\!\! \mathtt{TO\_UNITVECTOR}([\azi_k]_j,\![\ele_k]_j)$\,for\,$j\!\!=\!\!1,\!\ldots\!,n_s$
	\For{$i=1:\np$}
		\State $x_k^{(i)} \sim \N(A_{k-1}x_{k-1}^{(i)}, Q_{k-1})$
		\State $\widetilde{w}_k^{(i)} \gets \exp\!\left(\kappa \sum_{j=1}^{n_s} u_j^\t R_j \frac{[x_k^{(i)}]_\text{pos}-s_j}{\| [x_k^{(i)}]_\text{pos}-s_j \|}\right) \cdot w_{k-1}^{(i)}$
	\EndFor
	\State $w_k^{(i)} \gets \frac{\widetilde{w}_k^{(i)}}{\sum_{j=1}^{\np} \widetilde{w}_k^{(j)}}$ for $i\!=\!1,\ldots,\np$
	\State $x_{k|k} \gets \sum_{i=1}^{\np} w_k^{(i)} x_k^{(i)}$
	\If{$1/\sum_{i=1}^{\np} (w_k^{(i)})^2 < 0.1 \np$}
		\State $[x_k^{(1:\np)},w_k^{(1:\np)}] = \mathtt{RESAMPLE}(x_k^{(1:\np)},w_k^{(1:\np)})$
	\EndIf
\EndFor
\end{algorithmic}
\end{algorithm}
\begin{algorithm}[t]
\small
\caption{Measurement model function}\label{algo:meas_func}
\begin{algorithmic}[1]
\newcommand{\dl}{\delta}
\State \textbf{Inputs:} position $\pos$; anchor positions \& rotations $s_{1:n_s}$, $R_{1:n_s}$
\State \textbf{Outputs:} measurement model function value $c$ and Jacobian $C$
\For{$j=1:n_s$}
	\State $d_j \gets \frac{1}{\|\pos-s_j\|} (\pos-s_j)$
	\State $c_{3j-2:3j} \gets R_j d_j$
	\State $C_{3j-2:3j,\text{pos}} \gets \frac{1}{\|\pos-s_j\|} R_j \left( \eye_3 - d_j d_j^\t \right)$
	\State $C_{3j-2:3j,-\text{pos}} \gets \zeros$ \Comment{$-\text{pos}$: indices excluding $\text{pos}$ indices}
\EndFor
\end{algorithmic}
\end{algorithm}

\section{Simulations} \label{sec:tests}

We compare the proposed filters based on the \ac{vmf} and unit vector model \eqref{eq:vmf_model} with the filters based on the normal distribution and azimuth--elevation model \eqref{eq:normal_model}. The comparisons rely on numerical simulations computed with \textsc{Matlab}.

We study two different measurement models, from which the measurements are generated. In Model I the measurements are generated from the normal model \eqref{eq:normal_model} such that each direction has a unique azimuth--elevation representation; i.e.\ if the generated elevation measurement $[\ele_k]_j$ is negative (resp.\ larger than $\pi$), the elevation is flipped to its absolute value $|[\ele_k]_j|$ (resp.\ to the angle $2\pi\!-\![\ele_k]_j$) and the azimuth measurement $[\azi_k]_j$ is flipped to $[\azi_k]_j\!-\!\pi$. The flipping reflects a real-world equipment's behavior to have a unique representation of direction. In Model II the measurements are generated from the \ac{vmf} model \eqref{eq:vmf_model}.

\begin{figure}[t]
\centering
\includegraphics[width=0.48\columnwidth]{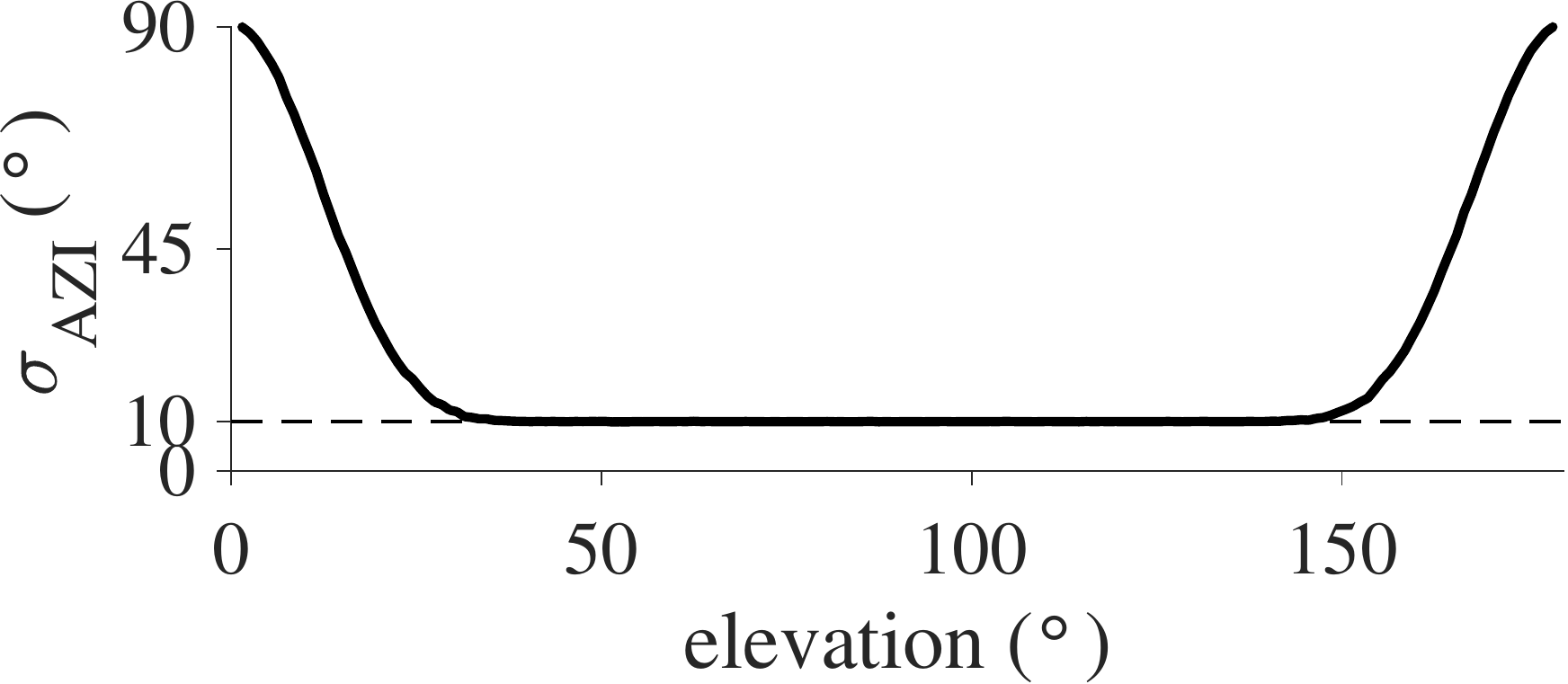}
\includegraphics[width=0.48\columnwidth]{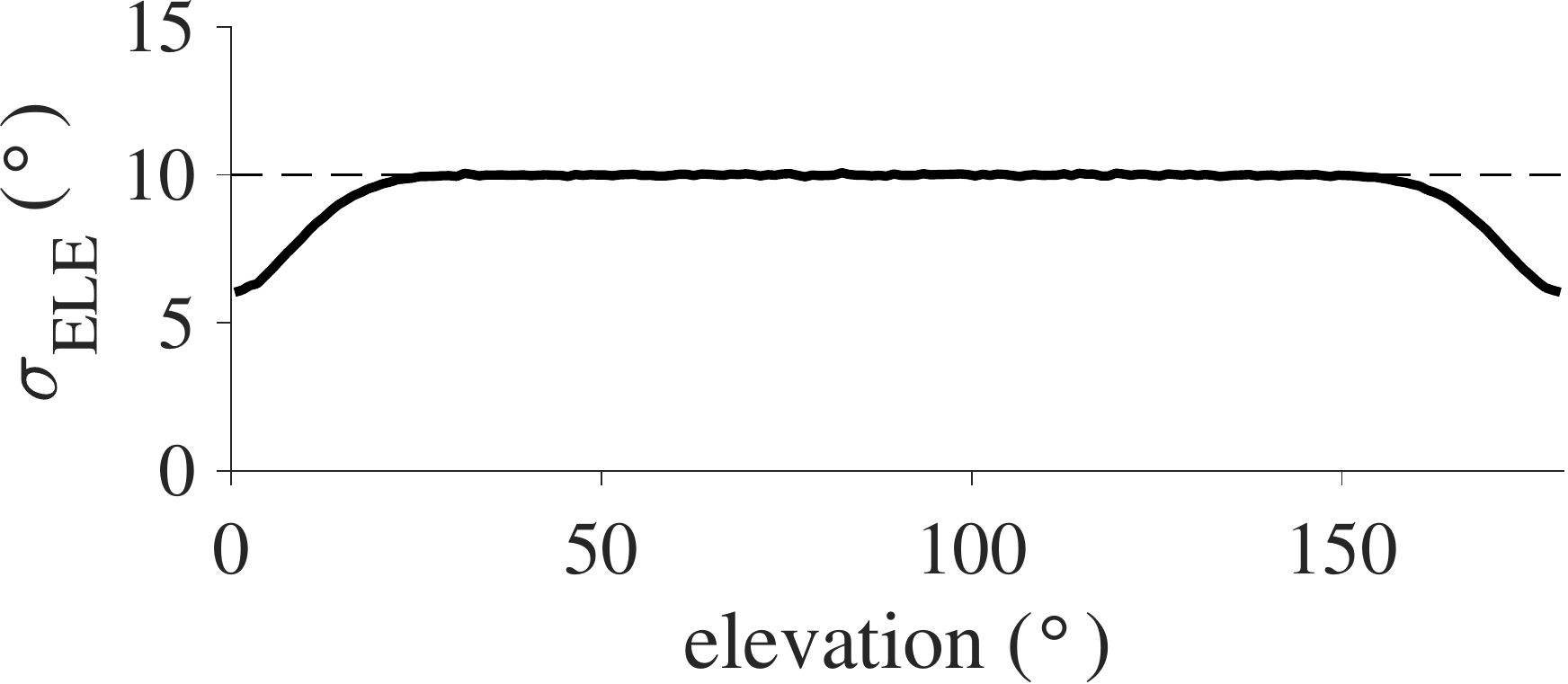}
\caption{Standard deviations of the azimuth (left) and elevation (right) measurement noises as a function of the elevation in the normal model with the flipping and $\sazi\!=\!\sele\!=\!10^\circ$.}
\label{fig:adaptivestds}
\end{figure}

We compare four different positioning filters:
\vspace{-3mm}
\begin{itemize}
\setlength\itemsep{-1.5mm}
\item AE-nominal: normal model \eqref{eq:normal_model}; with Model I $\sazi$ and $\sele$ are the scale parameters of the distribution from which the data were generated, with Model II $\sazi$ and $\sele$ are determined using \eqref{eq:sigma2kappa}.
\item AE-fitted: normal model \eqref{eq:normal_model}; $\sazi$ and $\sele$ fitted as the maximum likelihood parameters given $10^5$ simulated measurements generated for $10^5$ random directions.
\item AE-adaptive: normal model \eqref{eq:normal_model}; $\sazi$ and $\sele$ chosen as the standard deviations of the normal distribution with flipping at the given elevation; these standard deviations are pre-computed using a grid with 1-degree grid size for the elevation, and shown in Fig.\ \ref{fig:adaptivestds}.
\item VMF: \ac{vmf} model \eqref{eq:vmf_model}; with Model I $\kappa$ is determined using \eqref{eq:sigma2kappa}, with Model II $\kappa$ is the concentration parameter of the \ac{vmf} from which the data were generated.
\end{itemize}

\subsection{Model comparison} \label{sec:modelcomparison}

In this test we compute the expectations of the normal and \ac{vmf} log-likelihoods over the distribution $p(\azi,\ele,\pos)$. The conditional measurement distribution $p(\azi,\ele|\pos)$ is either the normal distribution with flipping (Model I) or the azimuth--elevation distribution implied by the \ac{vmf} distribution (Model II), and 3-dimensional position's distribution is the uniform distribution over the unit sphere $p(x)\!=\!\frac{1}{4\pi}$. The quantitative goodness of the normal fit is measured with
\begin{align} \label{eq:L_normal}
\mathcal{L}_{\N}
&= \int_0^\pi \!\!\!\!\int_{-\!\pi}^\pi \!\int_{\mathcal{S}_3(1)} \!\!\!\big[\log\N_{[-\pi,\pi]}( \azi\!-\!\atan_2(\pos_2,\pos_1); 0, \sazi^2) \nonumber\\
&+ \log\N_{[0,\pi]}(\ele\!-\!\atan_2(\pos_3,\|\pos_{1:2}\|); 0, \sele^2) \big] \nonumber\\
&\times p(\azi,\ele | \pos) \tfrac{1}{4\pi} \d\pos \d\azi \d\ele,
\end{align}
where $\mathcal{S}_3(1)$ is the 3-dimensional unit sphere, and $\N_\mathcal{A}$ is the normal distribution truncated to the set $\mathcal{A}$. The goodness of the \ac{vmf} fit is measured with the number
\begin{align} \label{eq:L_vmf}
\mathcal{L}_{\VMF} \!\!
=&\!\! \int_0^\pi \!\!\!\!\int_{-\!\pi}^\pi \!\!\int_{\mathcal{S}_3(1)} \!\!\!\!\!\!\!\big[ \log\!\VMF( \mathtt{TO\_UNITVECTOR}(\azi,\!\ele); \pos, \kappa ) \nonumber\\
+\log&(\sin(\ele\!+\!\tfrac{\pi}{2})) \big] \,p(\azi,\ele | \pos) \tfrac{1}{4\pi} \d\pos \d\azi \d\ele.
\end{align}
The term $\log(\sin(\ele\!+\!\tfrac{\pi}{2}))$ in \eqref{eq:L_vmf} comes from the transformation from Cartesian space's unit sphere into spherical coordinates' area with radius one. We computed \eqref{eq:L_normal} and \eqref{eq:L_vmf} using Monte Carlo integration with $10^5$ samples.

Table \ref{tab:model_comp} gives the used parameters as well as the model comparison numbers $\exp(\mathcal{L})$ normalised to sum to unity. The results show that \ac{vmf} explains the data better than AE-nominal and AE-fitted models for both Model I and Model II. AE-adaptive attempts to fix the problem of underestimating the azimuth's variance close to the poles, and has indeed the model comparison number close to the \ac{vmf} models especially for the normal model with flipping. The fitted maximum likelihood parameters of the AE model show large $\sazi$ compared to the nominal value because of the influence of the pole areas.
\begin{table}[t]
\small
\renewcommand{\arraystretch}{1.1}
\centering
\caption{Model comparison}
\label{tab:model_comp}
\begin{tabular}{c|cc|cc}
\multirow{2}{*}{Filter model} & \multicolumn{2}{c|}{Data from Model I} & \multicolumn{2}{c}{Data from Model II} \\
 & Parameters & $\exp(\mathcal{L})$ & Parameters & $\exp(\mathcal{L})$ \\
\hline
\multirow{2}{*}{AE-nominal} & $\sigma_\text{AZI}\!=\!$10 & \multirow{2}{*}{0.13} & $\sigma_\text{AZI}\!=\!$10 & \multirow{2}{*}{0.11} \\
& $\sigma_\text{ELE}\!=\!$10 && $\sigma_\text{ELE}\!=\!$10 & \\
\multirow{2}{*}{AE-fitted} & $\sigma_\text{AZI}\!=\!$18 & \multirow{2}{*}{0.21} & $\sigma_\text{AZI}\!=\!$21 & \multirow{2}{*}{0.24} \\
 & $\sigma_\text{ELE}\!=\!$10 && $\sigma_\text{ELE}\!=\!$10 & \\
AE-adaptive  && 0.33 && 0.30 \\
VMF-nominal & $\kappa\!=\!$33 & 0.32 & $\kappa=$33 & 0.35 \\
\end{tabular}
\end{table}

\subsection{A single measurement update example} \label{sec:singleupdate}

In this subsection we illustrate the difference of the azimuth--elevation and unit vector based filter updates. We use the prior distribution for the position $\pos \!\sim\! \N\left( \left[\begin{smallmatrix} 0.3\\-0.3\\-2 \end{smallmatrix}\right], 0.75^2 \eye_3 \right)$ and a direction measurement $\azi\!=\!-\pi$, $\ele\!=\!-\tfrac{19\pi}{20}$, $\sazi\!=\!\sele\!=\!5^\circ \tfrac{\pi}{180^\circ}$, which gives $\kappa\!=\!(\tfrac{180^\circ}{5^\circ \pi})^2$. The anchor is in the origin. The used \ac{ukf} parameter is $\lambda\!=\!0.5$, which provides equally weighted sigma points. We have intentionally placed the prior mean and the direction measurement close to and on different sides of the pole.

\begin{figure}[t]
\centering
\includegraphics[width=0.65\columnwidth,trim=5mm 5mm 5mm 20mm,clip=true]{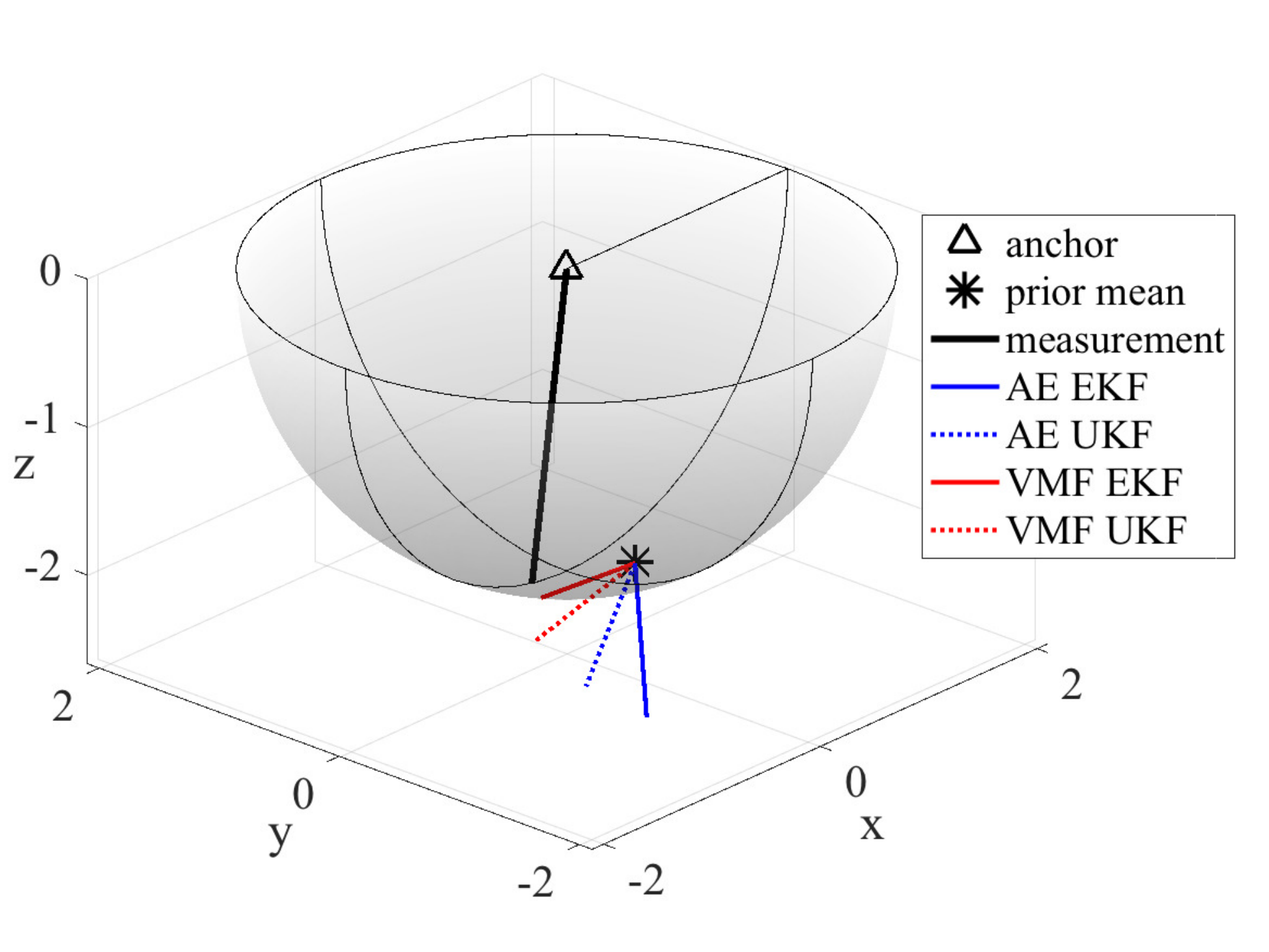}
\vspace{-5mm}
\caption{A single measurement update}
\label{fig:single_update}
\end{figure}

Fig.\ \ref{fig:single_update} shows the example scenario and the filter updates as line segments whose one end is in the prior mean and the other end is in the filtering posterior mean. The figure shows that the \ac{vmf} filter estimates are in directions close to the measurement direction, which is desirable because the prior distribution is quite diffuse and measurement is the only additional piece of information. The AE filters, on the contrary, update the estimate to an incorrect direction because the measurement model is highly nonlinear close to the pole. In \ac{ekf} the measurement function is linearised in the prior mean where the azimuth value has steepest descent direction in the clockwise tangential direction of the $xy$-plane's circle.

\subsection{Positioning tests with simulated data}

\begin{figure}[t]
\centering
\includegraphics[width=0.7\columnwidth]{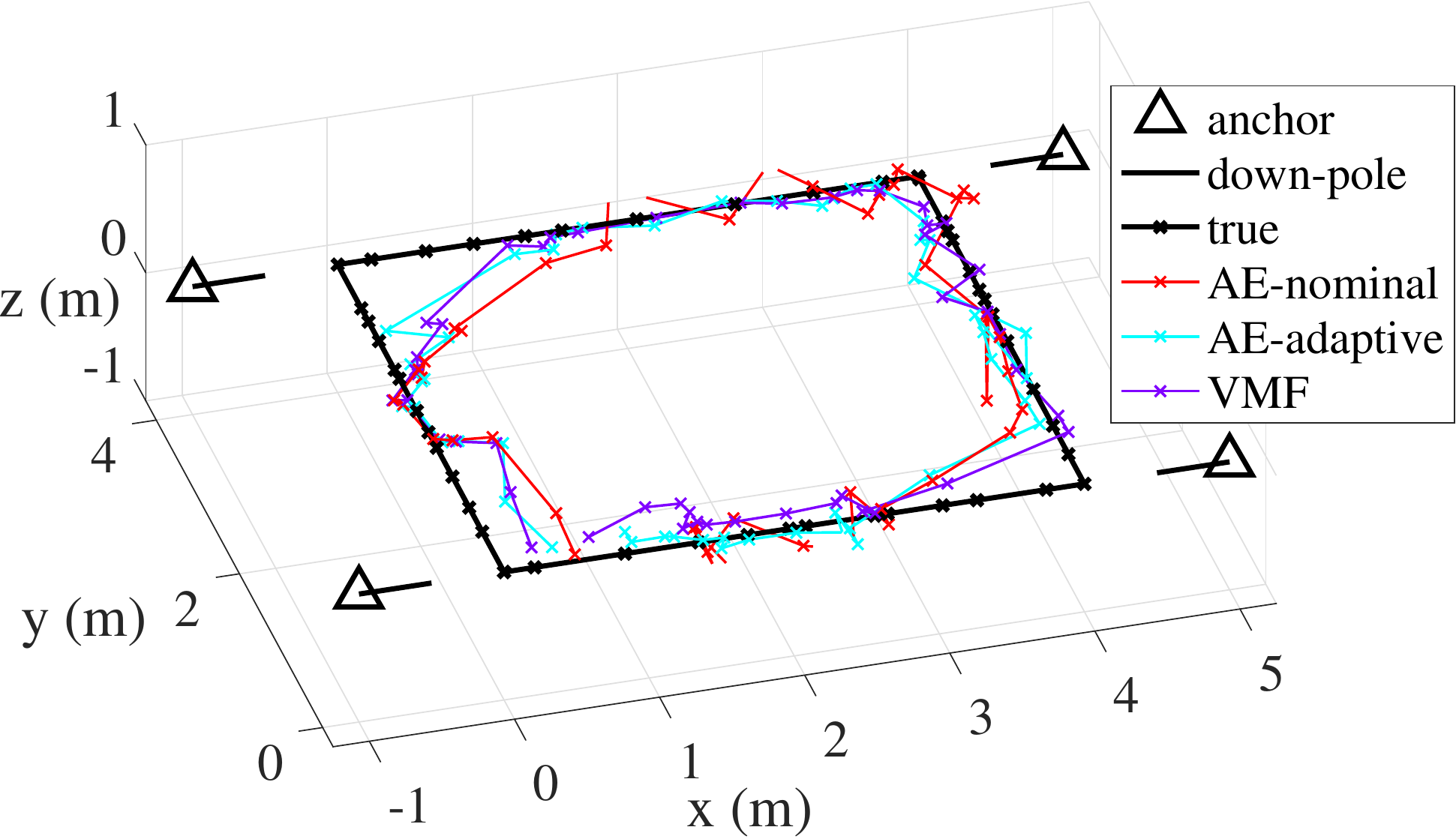}
\vspace{-4mm}
\caption{Test track and a simulated track with \ac{pf} estimates. The tracks starts in the origin and goes counter-clockwise.}
\label{fig:trackplot}
\end{figure}

We designed a $5\,\mathrm{m}\!\times\!5\,\mathrm{m}$ square-shaped test track in the horizontal $xy$-plane, and four anchors in the same plane. The test scenario was designed such that in two sides of the square the elevations of two anchors are zero. This is done to test how the algorithms perform close to the poles. We generated $10^4$ Monte Carlo replications of the test track and the \ac{aoa} measurements. The travelled distances between the measurement time instants were generated with the model
\begin{equation} \label{eq:simul_model}
\ell_k \sim \N\!\left(q_{xy} \sqrt{2\delta}\,\Gamma(1.5), \,2q_{xy}^2 \delta(1\!-\!\Gamma^2(1.5))\right),
\end{equation}
where $\Gamma$ is the Gamma function, the process noise parameter is $q_{xy}\!=\!0.5\,\mathrm{m}/\mathrm{s}^\frac{1}{2}$, and the time difference is $\delta\!=\!0.25\,\mathrm{s}$. The filter state includes only the user position, i.e.\ $x_k\!=\!\pos_k$, and the state-transition model is the random-walk model
\begin{equation}
\pos_k = \pos_{k-1} + w_{k-1},\quad w_{k-1}\sim\N\!\left(0_3,\delta\left[\begin{smallmatrix} q_{xy}^2 \eye_2&0_2\\0_2^\t&q_z^2 \end{smallmatrix}\right] \right),
\end{equation}
where $q_z$ is the process noise parameter for the $z$-direction for which we used $q_z\!=\!0.1\,\mathrm{m}/\mathrm{s}^\frac{1}{2}$. In this model, the random variable $\tfrac{1}{q_{xy}\sqrt{\delta}}\|[w_k]_{1:2}\|$ follows the chi-distribution with two degrees of freedom $\chi_2$ whose mean is $\sqrt{2}\Gamma(1.5)$ and variance $2(1\!-\!\Gamma^2(1.5))$, which gives the basis for the simulation model \eqref{eq:simul_model}. We used $10^4$ particles in \ac{pf} and the \ac{ukf} parameter value $\lambda\!=\!0.5$. Fig.\ \ref{fig:trackplot} shows an example of a simulated test track and the filter estimates. ``AE-adaptive'' filters use a non-additive measurement noise model where the measurement noise distribution's covariance matrix depends on the elevation through the formula explained in Section \ref{sec:modelcomparison}.

\begin{figure}[t]
\centering
\includegraphics[width=0.495\linewidth]{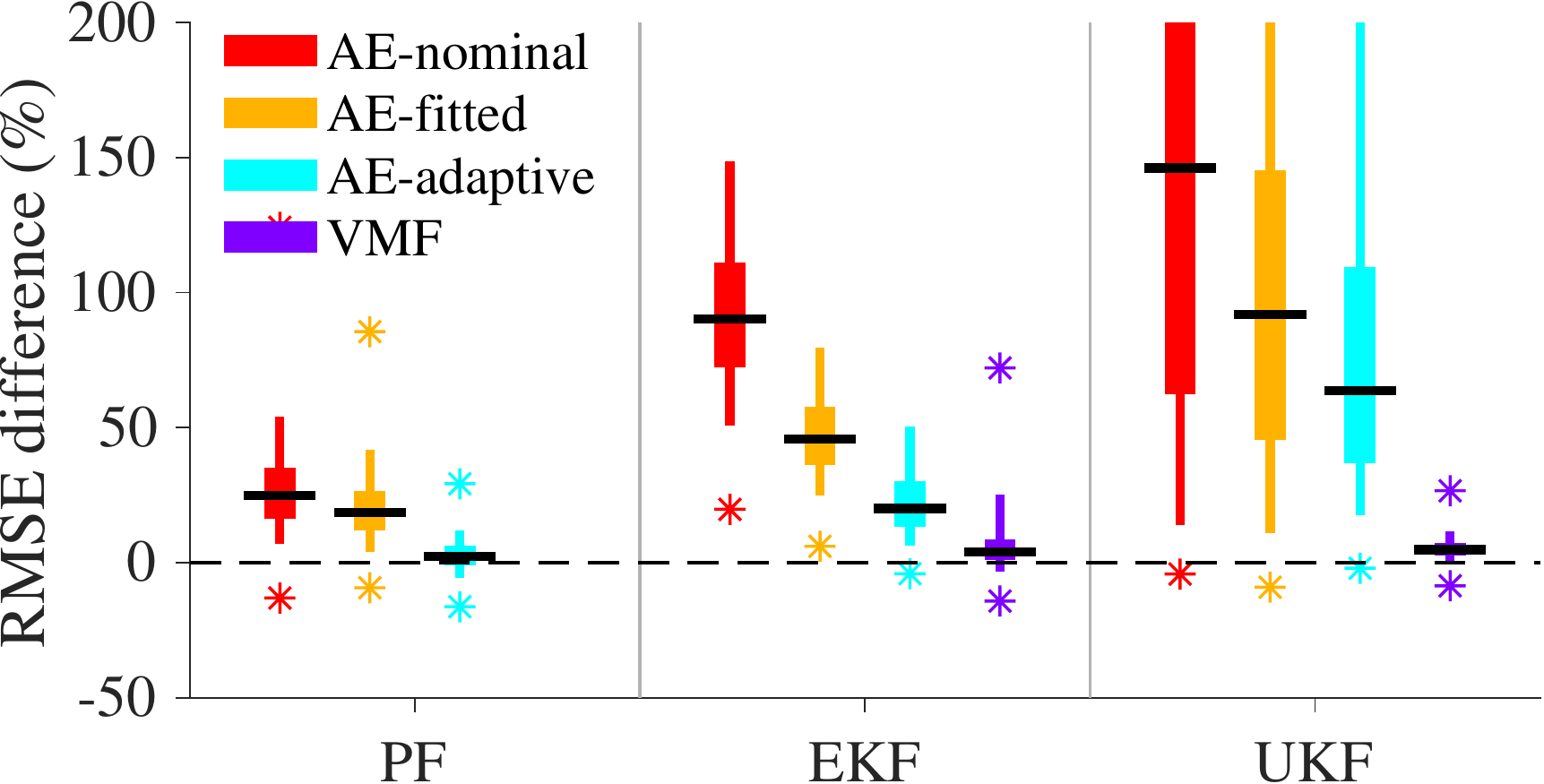}
\includegraphics[width=0.495\linewidth]{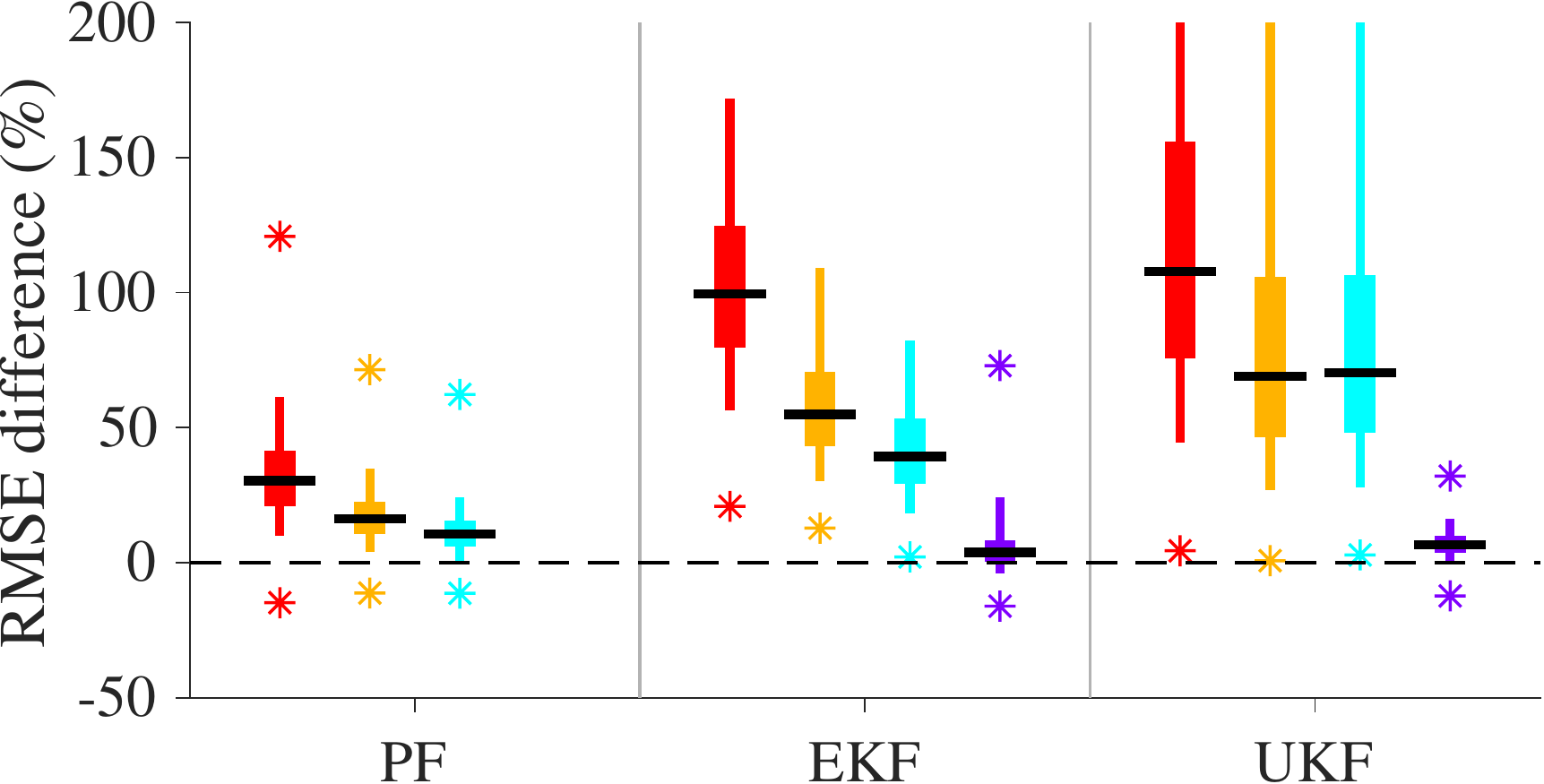}
\vspace{-7mm}
\caption{Percentual RMSE differences from PF-VMF. The measurement errors are generated from the normal distribution with flipping and $\sazi\!=\!\sele\!=\!10^\circ$ (left), and from the \ac{vmf} distribution with $\kappa=(\tfrac{180^\circ}{10^\circ\pi})^2$ (right).}
\label{fig:pos_tests_ae}
\end{figure}

Fig.\ \ref{fig:pos_tests_ae} shows the distributions of the percentual \ac{rmse} difference from the PF-VMF algorithm's \acp{rmse}. On the left, the measurement errors have been generated from Model I, and on the right from Model II. The box levels are 5\,\%, 25\,\%, 50\,\%, 75\,\%, and 95\,\% quantiles, and the asterisks show the minimum and maximum. The results show that the \ac{vmf} based algorithms greatly and systematically outperform the AE algorithms in accuracy. The differences are emphasised with \ac{ekf} and \ac{ukf}, which is probably due to the high nonlinearity of the measurement function close to the pole directions as explained in Section \ref{sec:singleupdate}. AE-adaptive filters are closer in accuracy to VMF than AE-nominal and AE-fitted, but in \ac{ekf} and \ac{ukf} the adaptivity does not necessarily improve the accuracy. This is probably due to the fact that these filters choose the measurement noise variance locally, in a single point in \ac{ekf} and in the sigma points in \ac{ukf}. Furthermore, \ac{vmf} based \ac{ekf} and \ac{ukf} are close to \ac{vmf} based \ac{pf} in accuracy.

\section{Conclusion} \acresetall \label{sec:conclusion}

In this article we propose modelling an \ac{aoa} positioning measurement as a \ac{vmf}-distributed unit vector instead of the conventional normally distributed azimuth and elevation measurements. Describing the 2-dimensional \ac{aoa} measurement with three numbers removes discontinuities and reduces nonlinearity at the poles of the azimuth--elevation coordinate system. Furthermore, the \ac{vmf} based model models the physical errors invariantly of rotations of the spherical coordinate system in which the azimuth and elevation measurements are expressed, which is sound if there is no reason to assume narrower and asymmetric error distributions in solid angle space close to the pole directions. The presented simulations show that when the user moves close to the pole directions, the proposed \ac{vmf} based \ac{pf}, \ac{ekf}, and \ac{ukf} algorithms show substantial improvement in the positioning accuracy.

\bibliographystyle{IEEEtran}
\bibliography{vmf}

\end{document}